\documentclass[letterpaper]{article}
\usepackage{aaai}
\usepackage{times}
\usepackage{helvet}
\usepackage{courier}
\usepackage{amsmath,amssymb,txfonts}
\usepackage{latexsym}
\usepackage{fancyhdr}
\usepackage{bigstrut}
\usepackage[dvipdfmx]{graphicx}
\usepackage{color}
\usepackage{url}
\AtBeginDvi{}

\begin{document}
%
\title{Collaboration on Social Media:\\ Analyzing Successful Projects on Social Coding}
\author{
Yuya Yoshikawa\\
Nara Institute of Science and Technology\\
8916-5, Takayama-cho, Ikoma-shi,\\
Nara, Japan\\
yuya-y@is.naist.jp
\And
Tomoharu Iwata\\
NTT Communication Science\\
Laboratories, NTT Corporation\\
2-4, Hikaridai, Seika-cho,\\
Soraku-gun, Kyoto, Japan\\
iwata.tomoharu@lab.ntt.co.jp
\And
Hiroshi Sawada\\
NTT Service Evolution Laboratories,\\
NTT Corporation\\
1-1 Hikari-no-Oka, Yokosuka-shi,\\
Kanagawa, Japan\\
sawada.hiroshi@lab.ntt.co.jp
}
\maketitle
%
%
%
%
\begin{abstract}
\begin{quote}
Social Coding Sites (SCSs) are social media services for sharing software development projects on the Web, and many open source projects are currently being developed on SCSs.
One of the characteristics of SCSs is that they provide a platform on social networks that encourages collaboration between developers with the same interests and purpose. 
For example, external developers can easily report bugs and improvements to the project members.

In this paper, we investigate keys to the success of projects on SCSs based on large data consisting of more than three hundred thousand projects.
We focus on the following three perspectives: 1) the team structure, 2) social activity with external developers, and 3) content developed by the project.
To evaluate the success quantitatively, we define activity, popularity and sociality as success indexes.
A summary of the findings we obtained by using the techniques of correlation analysis, social network analysis and topic extraction is as follows:
the number of project members and the connectivity between the members are positively correlated with success indexes.
Second, projects that faithfully tackle change requests from external developers are more likely to be successful.
Third, the success indexes differ between topics of softwares developed by projects.
Our analysis suggests how to be successful in various projects, not limited to social coding.
\end{quote}
\end{abstract}

%
%
%
%
\section{Introduction}
Social Coding Sites (SCSs) are social media services for sharing software development projects on the Web. 
Typical SCSs are GitHub\footnote{\url{https://github.com/}} and Bitbucket\footnote{\url{https://bitbucket.org/}}.
These SCSs provide environments for development and visualize developers' activities and the history of software changes on the Web.
The number of SCS users including organizations are rapidly increasing, and GitHub and Bitbucket currently have three millions and one million users, respectively.

On SCSs, developers are free to start projects and develop a software for each project.
The development can be undertaken by collaborating with multiple developers.
One of the characteristics of the SCSs is that they provide a platform that encourages collaboration between developers with the same interests and purpose. 
For example, external developers can easily report bugs and improvements or request the project members to change the software.
This is highly beneficial in terms of opening projects on SCSs to the public.
The SCSs have a social network of developers as with various social media sites, and developers' activities such as project making, contribution to projects and bookmarks are shared through the social network.
The social network is useful for finding projects and producing new collaborations.


What are the differences between successful and unsuccessful projects on the SCSs?
In this paper, we aim to investigate the characteristics of successful social coding projects.
To evaluate the success quantitatively, we define the following three measures as success indexes: activity, popularity and sociality.
The activity is calculated based on the update frequency of content developed by each project.
The popularity and sociality are evaluated by the number of bookmarks and the number of updates from the external developers. 
We perform an analysis using the success indexes by focusing on three perspectives: 1) the team structure, 2) social activity with external developers, and 3) content developed by the project.
To investigate the relationships between team structure and success, we first construct a collaboration network for each project, and perform a correlation analysis using measurements based on social network analysis.
Furthermore, we investigate the efficient number of members and the workload bias among members.
From the perspective of social activity, we investigate the effects of the response to the external developers, that is, the reaction of the internal developers of a project when bugs and improvements are reported by the external developers.
Finally, we investigate what successful projects develop based on README files.
To extract the software characteristics from README, we employ a topic model.
Through this analysis, we show that the success indexes differ among software topics.

\begin{figure}[t]
\begin{center}
\includegraphics[width=87mm]{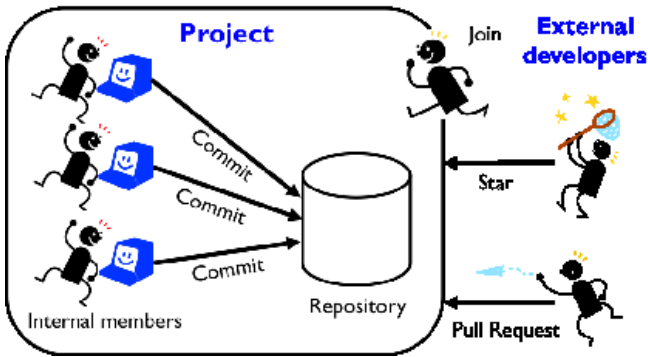}
\end{center}
\caption{The development process and developer behavior on social coding.}
\label{fig:social_coding}
\end{figure}

To best of our knowledge, our study is the first to analyze the characteristics of successful social coding projects from various perspectives.
Our analysis suggests how to achieve success in various projects, not only projects on social coding.
One reason is that the coverage is wide since we treat more than three hundred thousand projects as subjects for analysis.
Also, the types of projects vary from personal projects to enterprise projects.
By thoroughly investigating data collected from the SCSs, we believe that we can discover the law of success for the projects.

The rest of the paper is organized as follows.
In the second section, we describe social coding and our datasets, and define project success indexes.
In the third section, we perform a basic analysis of the dataset to understand the world of social coding.
In sections four to six, we investigate the characteristics of successful projects from the above three perspectives.
In the seventh section, we review related work.
The last section concludes this paper with knowledge obtained from our analysis.

\begin{figure}[t]
\begin{center}
\includegraphics[width=85mm]{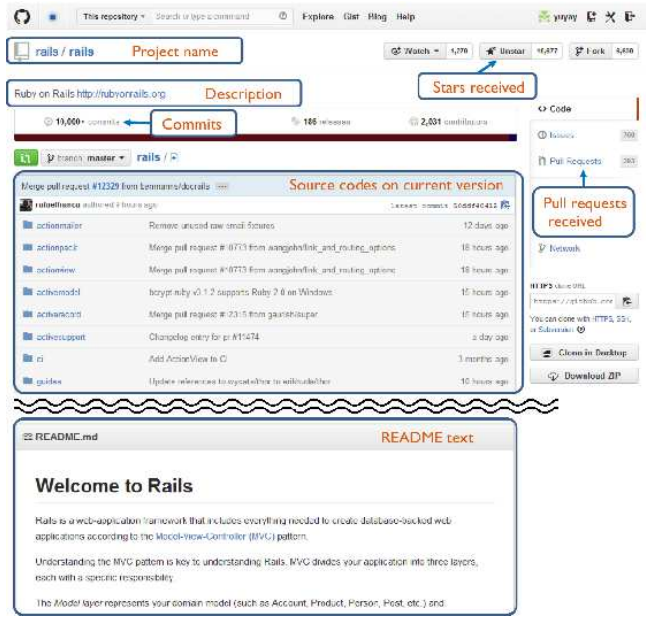}
\end{center}
\caption{Screen-shot of the front page of the Ruby on Rails project on GitHub.}
\label{fig:github}
\end{figure}

%
%
%
%
\section{Social Coding}\label{sec:social_coding}
In this section, we introduce a development process and developer behavior on social coding.
Then, we define three success indexes with which to measure project success.

%
%
\subsection{What Is Social Coding?}
We introduce developer behavior from two perspectives, inside and outside a project.
Figure \ref{fig:social_coding} shows a simplified schematic of the development process and the developer behavior with respect to social coding.
First, we introduce the behavior inside a project.
In projects on social coding, there are one or more developers as {\it internal members} of the project.
Each project has a {\it repository} consisting of the source codes and software documents. 
In each project, the internal members of the project write new codes and documents or change the existing ones to improve the software or fix bugs, and add them to the repository via an operation called {\it commit}. 
Next, we introduce the behavior outside a project.
The developers outside the project can bookmark the project if they think the software created by the project is excellent. 
This operation is also called labeling with a {\it star}.
The developers outside a project can join the project by sending {\it pull requests} with codes that they wrote.
Although whether or not the codes are accepted to the project is determined by the internal members, the pull requests enable the internal and external members to collaborate.

Figure~\ref{fig:github} shows a screen-shot of the front page of the Ruby on Rails project, which is a typical project on GitHub.
On this page, we can understand who is managing the project, what the project is and how it is managed by seeing a brief description, source codes and the contents of README.
In addition, the number of commits, and the number of stars and pending pull requests that have been received are displayed.

%
%
\begin{table}[t]
\caption{Specifications of collected GitHub dataset.}
\centering
\begin{tabular}{rr}
\hline
\ & Number \\
\hline
Number of projects & 317,077 \\
Total commits & 41,720,139 \\
Total stars & 5,164,934 \\
Total pull requests & 1,903,595 \\
Number of developers & 1,381,121 \\
Total followings & 1,885,266 \\
\hline
\end{tabular}
\label{tab:data_stats}
\end{table}

\begin{figure*}[t]
  \begin{center}
  \begin{minipage}{0.32\hsize}
    \begin{center}
      \includegraphics[width=60mm]{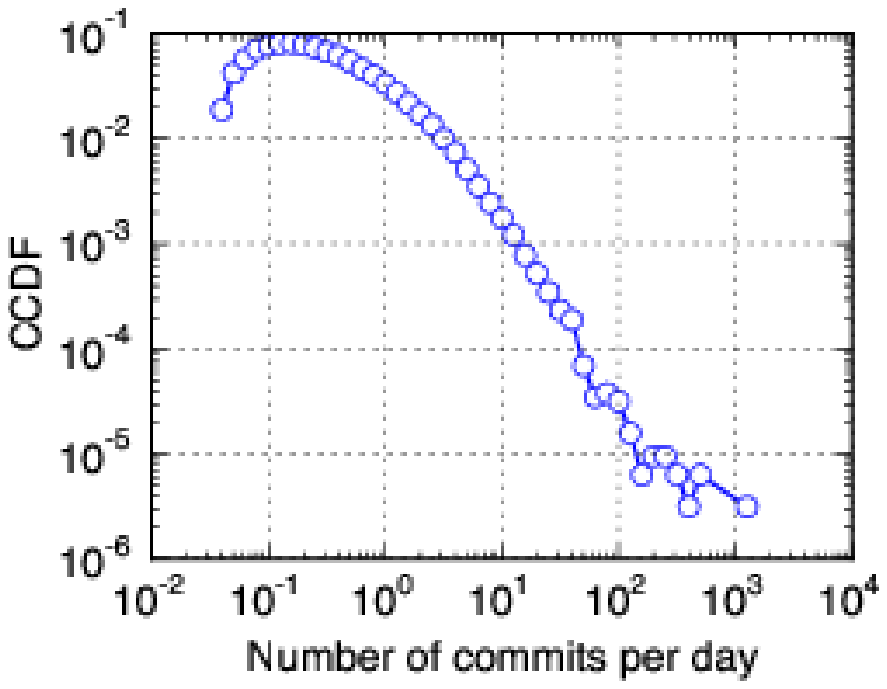}\\
      \vspace{-2mm}      
      (a) $\mathtt{Commit}$
    \end{center}
  \end{minipage}
  \begin{minipage}{0.32\hsize}
    \begin{center}
      \includegraphics[width=60mm]{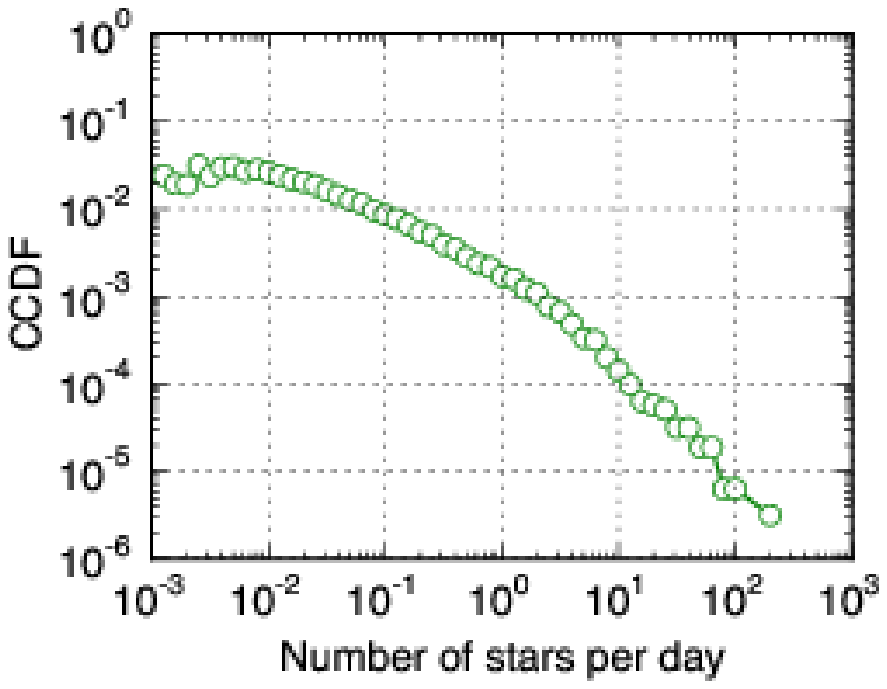}\\
      \vspace{-2mm}      
      (b) $\mathtt{Star}$
    \end{center}
  \end{minipage}
  \begin{minipage}{0.32\hsize}
    \begin{center}
      \includegraphics[width=60mm]{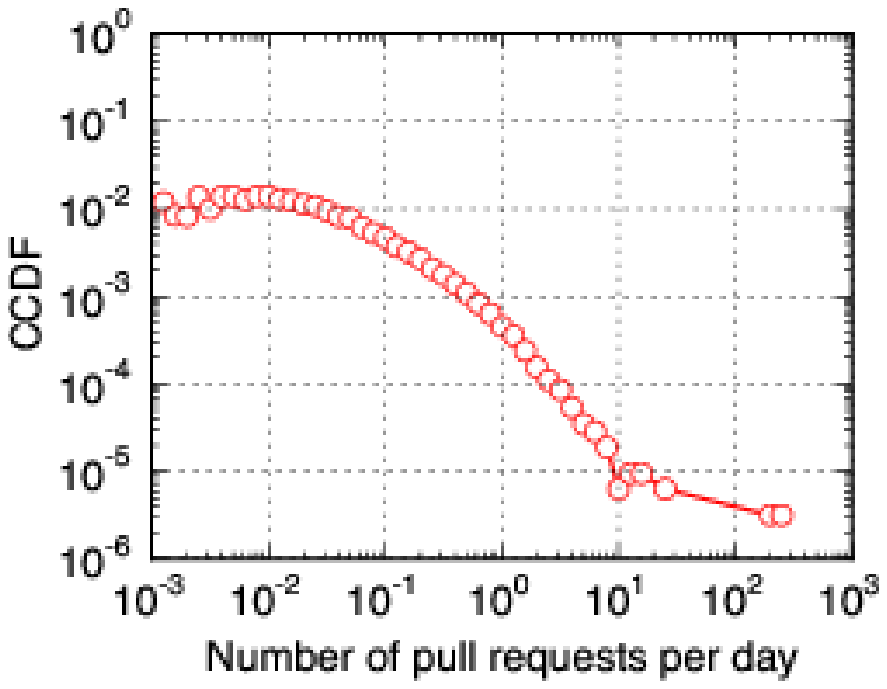}\\
      \vspace{-2mm}      
      (c) $\mathtt{PullReq}$
    \end{center}
  \end{minipage}
 \caption{Distributions of project success indexes. The vertical axes indicate complementary cumulative distribution function (CCDF).}
 \label{fig:success_dist}
 \end{center}
\end{figure*}

\begin{figure*}[t]
  \begin{center}
  \begin{minipage}{0.31\hsize}
    \begin{center}
      \includegraphics[width=60mm]{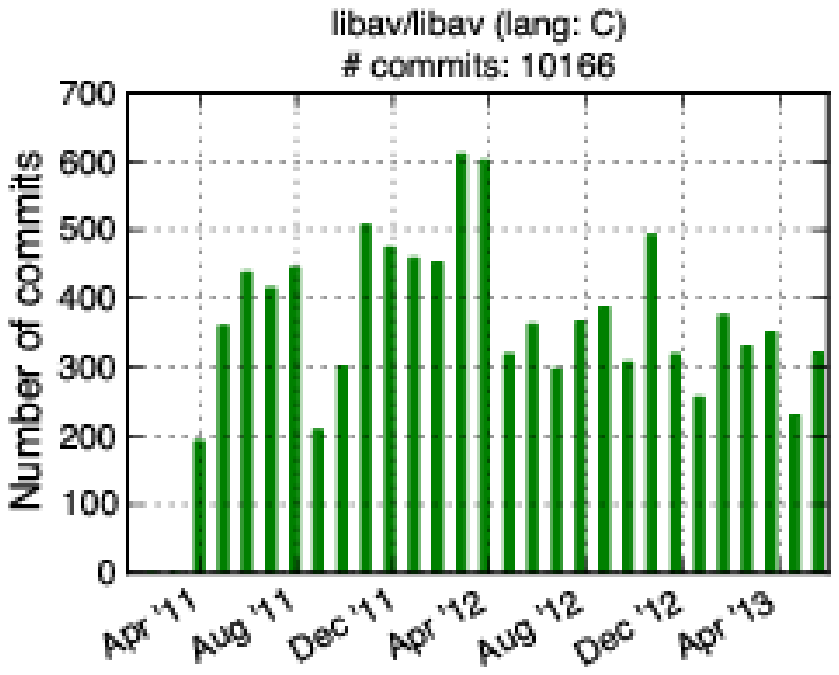}\\
      \vspace{-2mm}      
      (a) Uniform Update (UU)
    \end{center}
  \end{minipage}
  \begin{minipage}{0.32\hsize}
    \begin{center}
      \includegraphics[width=60mm]{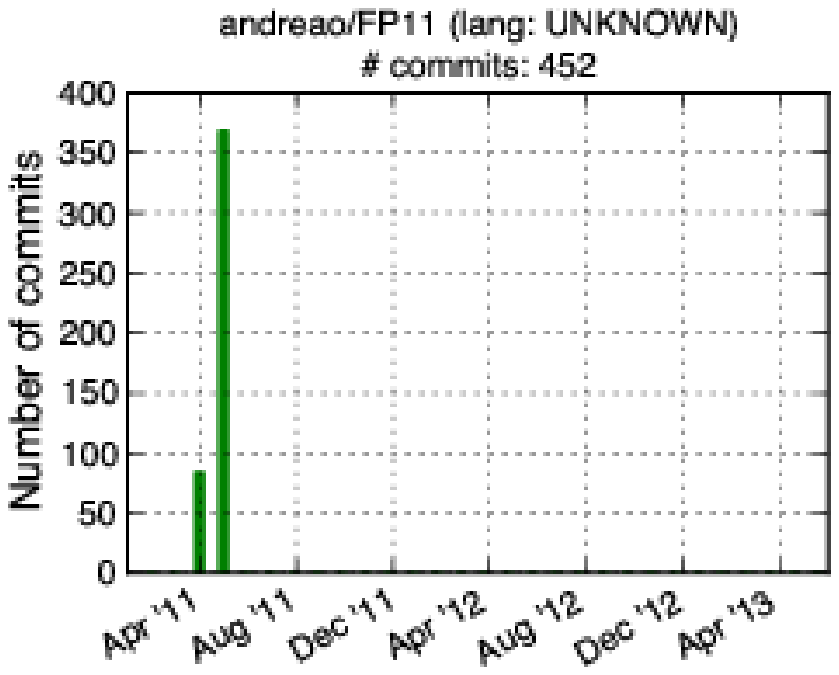}\\
      \vspace{-2mm}      
      (b) Single Concentrated Update (SCU)
    \end{center}
  \end{minipage}
  \begin{minipage}{0.35\hsize}
    \begin{center}
      \includegraphics[width=60mm]{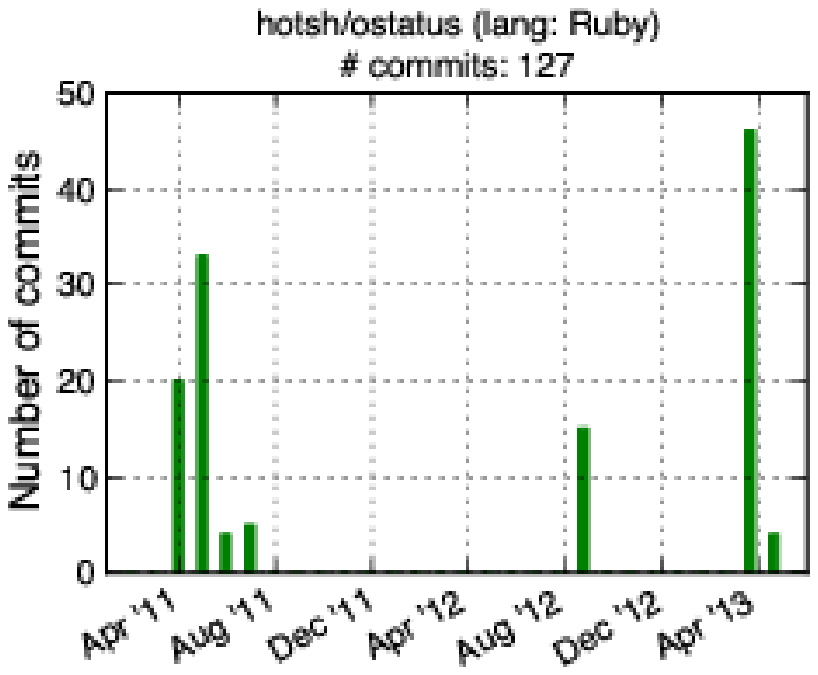}\\
      \vspace{-2mm}      
      (c) Multiple Concentrated Update (MCU)
    \end{center}
  \end{minipage}
 \caption{Examples of growth patterns of projects.}
 \label{fig:commit_trends}
 \end{center}
\end{figure*}

\subsection{GitHub Dataset}
In this paper, we use data obtained from Github, which is the most typical social coding site, through Github Archive\footnote{\url{http://www.githubarchive.org}} and Github API\footnote{\url{http://developer.github.com}}.
Github Archive provides 18 kinds of public events such as commits, pull requests and stars that have occurred since February 2011.
We crawled all the events until May 2013.
We then extracted projects for which there were more than 30 commits and which are not copy projects.
For the extracted projects, we collected information about internal members, commits, pull requests and stars.
Additionally, we crawled the README files of the projects using Github API to obtain texts representing the overviews of the projects.

In Github, developers can follow the activities of their favorite developers using {\it follow} function similar to that in Twitter.
We also collected the information of developers who did more than a commit or a following, and constructed a follow network by placing a link from developer $i$ to developer $j$ if $i$ followed $j$.
Table \ref{tab:data_stats} shows the specifications of the collected dataset.

%
%
\subsection{Project Success indexes}
%

To quantify the success of a project, we define the following three numbers as {\it success indexes}.

\vspace{2mm}
\noindent{\bf Number of commits} ($\mathtt{Commit}$) represents the number of times codes and documents are added to the repository of the project.
Projects with higher $\mathtt{Commit}$ are more likely to develop quickly and effectively.
We use it for measuring the activity of a project.

\vspace{2mm}
\noindent{\bf Number of stars} ($\mathtt{Star}$) represents the number of developers who marked the project with a star.
Stars are generally used to monitor the behavior of a project or watch the project later.
We use it to measure the popularity of a project.

\vspace{2mm}
\noindent{\bf Number of pull requests} ($\mathtt{PullReq}$) represents the number of times the external developers requested that their codes be accepted.
Since a higher $\mathtt{PullReq}$ indicates that the project is frequently discussed with external developers, we use it to indicate the sociality of a project.

\vspace{2mm}
\noindent
We note that these indexes are normalized by the number of days a project has continued.
Below we investigate project structures and behaviors which are related to the difference of the success indexes.

%
%
%
%
\section{Basic Analysis}
Before beginning our analysis in terms of the success, we illustrate the characteristics of projects and developers' relationships to understand the world of social coding based on the following questions:
\begin{enumerate}
\item[(1)] How often are projects updated?
\item[(2)] How often are the projects watched by external developers?
\item[(3)] How do the projects grow?
\item[(4)] How are the developers connected with each other? 
\end{enumerate}
%
%
\subsection{Statistics of Projects}
We answer questions (1) and (2) by analyzing the $\mathtt{Commit}$, $\mathtt{Star}$ and $\mathtt{PullReq}$ statistics.
Figure~\ref{fig:success_dist} shows the distributions of the project success indexes.
As shown in these figures, we find these indexes are biased toward small values.
In fact, while the mean values of $\mathtt{Commit}$, $\mathtt{Star}$ and $\mathtt{PullReq}$ for the projects are 0.62, 0.05 and 0.02, their median values are 0.22, 0.00 and 0.00.
Furthermore, 53\% of the projects are zero $\mathtt{Star}$ and 70\% of the projects are zero $\mathtt{PullReq}$.
According to these results, many projects would not be found or evaluated by the external developers of projects.

%
%
\subsection{Growth Patterns of Projects}
To answer question (3), we show how each project has grown by counting the number of commits for each month. 
By looking over the growth patterns of the projects, we found that there are three types of growth patterns.
%
\begin{itemize}
\item {\bf Uniform Update (UU)}:
this indicates projects in which developers almost constantly added or changed codes.
\item {\bf Single Concentrated Update (SCU)}:
this indicates projects that stagnated within a few months of their starting day.
\item {\bf Multiple Concentrated Update (MCU)}:
this indicates projects with multiple periods in which many commits occurred, although there are no commit in most months.
\end{itemize}
\begin{figure}[t]
\begin{center}
\includegraphics[width=65mm]{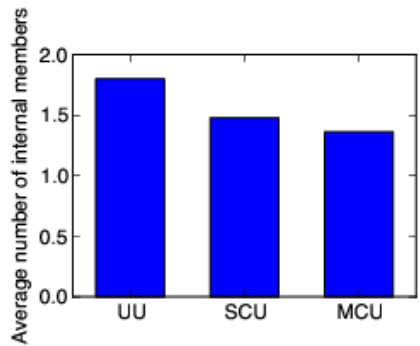}
\end{center}
\vspace{-6mm}
\caption{Average number of internal members involved in projects with each growth pattern.}
\label{fig:pattern_vs_member}
\end{figure}

Figure~\ref{fig:commit_trends} shows examples of the growth patterns.
We classified 200 randomly sampled projects into these patterns by hand.
The percentages of these patterns were 10\% (Uniform Update), 45\% (Single concentrated update) and 45\% (Multiple concentrated update).
Figure~\ref{fig:pattern_vs_member} shows the difference between the average numbers of internal members from projects with each pattern. 
We find that UU pattern projects have many more internal members than the other two patterns.

%
\begin{figure}[t]
\begin{center}
\includegraphics[width=65mm]{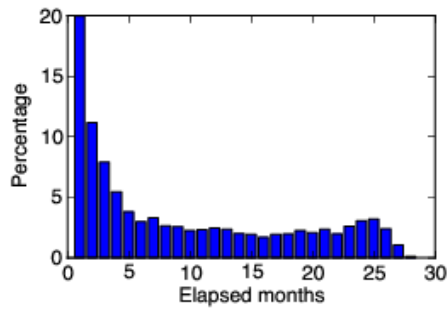}
\end{center}
\vspace{-6mm}
\caption{Distribution of the number of months taken to exceed 90\% total commits.}
\label{fig:over90percent}
\end{figure}

We next analyze how long projects last.
Since we cannot know when the projects terminated, we instead measure the number of months taken to exceed 90\% total commits for each project.
Figure \ref{fig:over90percent} shows the distribution of the elapsed months.
We found that about 20\% of the projects terminate within a month of the project starting day and 
about 40\% of the projects terminate within three months of the project starting day.
On the other hand, we also found that about 20\% of projects terminate after continuing for more than 20 months.

%
%
\subsection{Structure of Follow Network}
At the end of this section, we analyze the properties of the connections between developers based on the follow network including all the nodes to answer question (4).

\begin{figure}[t]
\begin{center}
\includegraphics[width=65mm]{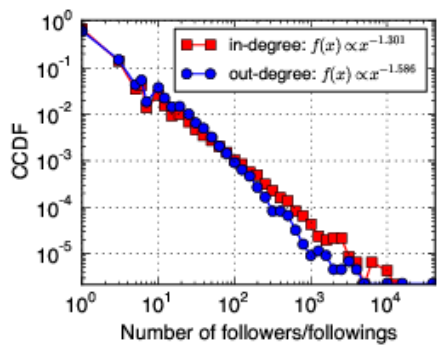}
\end{center}
\vspace{-6mm}
\caption{Degree distributions of follow network. Red squares represent in-degree (number of followers) distribution and blue circles represent out-degree (number of followings) distribution. }
\label{fig:follow_dist}
\end{figure}

First, we analyze degree distributions of the follow network.
Figure \ref{fig:follow_dist} shows the in-degree and out-degree distributions.
The vertical axis indicates complementary cumulative distribution function (CCDF), which is the normalized frequency of the degrees.
It is commonly known that the degree distributions of social networks obey a power-law distribution $f(x) \propto x^{-\alpha}$ \cite{Clauset2009}, where $\alpha$ is a parameter of the power-law distribution.
In the follow network of Github, we find $\alpha = 1.301$ for the in-degree and $\alpha = 1.586$ for the out-degree.
Since previous studies reported $\alpha = 2.276$ on Twitter~\cite{Kwak2010} and $\alpha = 2 \sim 3$ on general social networks~\cite{Clauset2009}, we can say that GitHub has relatively small $\alpha$ values. 
One of the reasons for small in-degree parameter values would be that there are no famous people who most developers know unlike Twitter.

We find that the CCDF of the in-degree distribution is bigger than that of the out-degree distribution at $x > 10^2$.
There are generally limitations to follow many developers by hand.
On the other hand, developers with many followers can easily arise thanks to a mechanism called "The rich get richer."
Such a phenomenon is also present in Twitter \cite{Kwak2010}.

Second, we analyze reciprocity to understand the property of the relationships between developers.
Reciprocity is the ratio of bidirectional edges on a directed network.
We found that the reciprocity on GitHub is 19.1\%, which is smaller than the 22.1\% found with Twitter \cite{Kwak2010}.
One reason is that developers follow others to collect information rather than for communication.



\begin{figure*}[t]
  \begin{center}
  \begin{minipage}{0.45\hsize}
    \begin{center}
      \includegraphics[width=50mm]{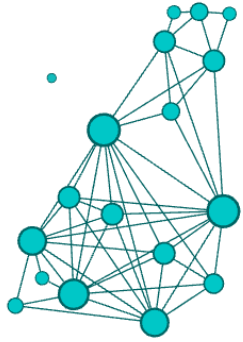}\\
      \vspace{2mm}      
      (a) high $\mathtt{Commit}$ project ($\mathtt{Commit} = 12.71$)
    \end{center}
  \end{minipage}
  \begin{minipage}{0.45\hsize}
    \begin{center}
      \includegraphics[width=50mm]{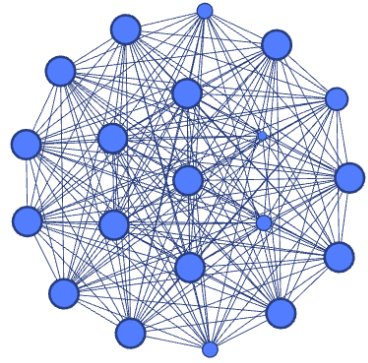}\\
      \vspace{1mm}      
      (b) low $\mathtt{Commit}$ project ($\mathtt{Commit} = 0.13$)
    \end{center}
  \end{minipage}
 \vspace{1mm} 
 \caption{Examples of collaboration networks in successful and unsuccessful projects. The node size indicates the number of edges that the node have.}
 \label{fig:collaboration_networks}
 \end{center}
\end{figure*}

%
%
%
%
\section{Effect of Team Structure}\label{sec:member}
In this section we analyze the effect of the team structure of a project on its success.
Here we focus on internal members of projects to investigate the team structure, that is, we do not consider the external developers that contribute to the projects by sending pull requests.
This is because we restrict our research object to a range that the project owner can control.

To understand the team structures, we construct a collaboration network for each project to represent the relationships between internal members in a development.
Each collaboration network is constructed by positioning an edge between internal members who join the same projects other than the given project at least once. 
The networks hereby describe how often the pairs of internal members work together.

\begin{table}[t]
\caption{Features representing team structure.}
\vspace{2mm}
\centering
\begin{tabular}{rp{17em}}
\hline
Feature & Description \\
\hline
$\mathtt{NumMember}$ & Number of internal members in a project. \\
$\mathtt{EdgeDense}$ & Number of edges on a collaboration network divided by the number of possible direct edges. \\
$\mathtt{ClustCoef}$ & Clustering coefficient of a collaboration network. \\
$\mathtt{CompRate}$ & Ratio of the number of connected components to the number of nodes. \\
$\mathtt{PathLen}$ & Average shortest path length of a collaboration network. \\
\hline
\end{tabular}
\label{tab:member_features}
\end{table}

We extract the features of the networks based on methods of social network analysis.
Table \ref{tab:member_features} lists the features and their descriptions.

%
%
\subsection{Correlation Analysis}

\begin{table}[t]
  \centering
  \caption{Kendall rank correlation coefficient between the team structure and the success of a project. All the elements exhibit statistically significant correlations (t-test, $p < 0.01$).}
  \vspace{2mm}
    \begin{tabular}{r|ccc}
    \hline
          & $\mathtt{Commit}$ & $\mathtt{Star}$ & $\mathtt{PullReq}$ \\
    \hline
    $\mathtt{NumMember}$ & 0.16  & 0.11  & 0.18 \\
    $\mathtt{EdgeDense}$ & 0.09  & 0.10  & 0.16 \\
    $\mathtt{ClustCoef}$ & -0.04 & 0.08 & 0.13 \\
    $\mathtt{CompRate}$ & -0.09 & -0.11 & -0.17 \\
    $\mathtt{PathLen}$ & 0.11 & 0.15 & 0.16 \\
    \hline
    \end{tabular}%
  \label{tab:member_corr}%
\end{table}%
To understand the relationships between team structure and the success of a project quantitatively, we first perform a correlation analysis based on the Kendall rank correlation coefficient (indicated by $\tau$), which is known as a robust index to outliers.
Table \ref{tab:member_corr} shows the Kendall rank correlation coefficients between the characteristics representing the team structure and the success indexes.
We first see the relationship between team size and success.
As shown in the table, we can see that $\mathtt{NumMember}$ correlates with the success indexes ($\tau = 0.11 \sim 0.18$). 
Although it is natural that $\mathtt{Commit}$ increases with increasing $\mathtt{NumMember}$, it is interesting that $\mathtt{Star}$ and $\mathtt{PullReq}$, which are evaluations from external developers, have relationships with $\mathtt{NumMember}$.
This would be because the sum of the influences of the internal members affects $\mathtt{Star}$ and $\mathtt{PullReq}$.

Second, we see the relationship between the connectivity of internal members and success.
$\mathtt{EdgeDense}$ represents how well the internal members connect in each project.
As shown in the table, we find that $\mathtt{EdgeDense}$ is positively correlated with all the success indexes, particularly $\mathtt{PullReq}$. ($\tau = 0.09 \sim 0.16$).
$\mathtt{CompRate}$ represents how the connections between internal members are divided.
We can explain this result from the fact that $\mathtt{CompRate}$ is negatively correlated with all the success indexes ($\tau = -0.09 \sim -0.17$).
According to those results, successful projects are more likely to exhibit strong connectivity between the internal members.

It is interesting that $\mathtt{PathLen}$ is positively correlated with all the success indexes ($\tau = 0.11 \sim 0.16$) because high $\mathtt{EdgeDense}$ tends to correspond to low $\mathtt{PathLen}$ ($\tau = -0.34$ between $\mathtt{EdgeDense}$ and $\mathtt{PathLen}$).
The average path length of a network is the smallest when the network is {\it complete}, which is that each node connects all other nodes \cite{Watts1988}.
Figure~\ref{fig:collaboration_networks} shows examples of collaboration networks in successful and unsuccessful projects in terms of $\mathtt{Commit}$.
Here these networks have similar $\mathtt{NumMember}$, relatively large $\mathtt{EdgeDense}$ and $\mathtt{ClustCoef}$, and relatively small $\mathtt{CompRate}$ in all projects.
We find that the network of Figure~\ref{fig:collaboration_networks}(a) is connected through some hub nodes except a isolated node.
However, the network is not complete, and its $\mathtt{PathLen}$ is 1.91.
On the other hand, the network of Figure~\ref{fig:collaboration_networks}(b) is almost complete, and its $\mathtt{PathLen}$ is 1.03.
This means that the project consists of members who collaborate with each other in other projects.
Thus, it is important to make a team such that some members know most other members and the members play a role connecting between all the members, rather than all members know all the other members. 

%

%
%
\subsection{Efficient Number of Internal Members}
As mentioned above, the team size of a project ($\mathtt{NumMember}$) is positively correlated with the success indexes.
In projects undertaken by organizations, it would be desirable to reduce the number of internal members due to budget limitations.
So, is there the optimal team size for success?
In answer to this question, Hoegl said, {\it Research evidence does not provide an absolute optimal team size in terms of a specific number, nor is there any conclusive indication of an absolute optimal range}~\cite{Hoegl2005}.
This is because the optimal number of team members would depend on the work to be performed.
On the other hand, some studies argued that small teams tend to produce better quality teamwork~\cite{Hoegl2005,Mueller2012}.
We then ask a question:
how does the team size affect teamwork quality in social coding?
To answer this question, we investigate the relationships between the team size and the number of commits of each member to reveal the teamwork quality.

%
\begin{figure}[t]
\begin{center}
\includegraphics[width=83mm]{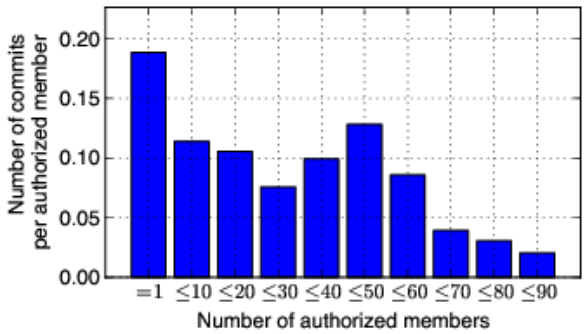}
\end{center}
\vspace{-6mm}
\caption{The relationship between the number of internal members and $\mathtt{Commit}$ divided by the number of internal members.}
\label{fig:member_vs_commits}
\end{figure}

Figure~\ref{fig:member_vs_commits} shows the relationship between the number of internal members and $\mathtt{Commit}$ divided by the number of internal members.
We first find that a one person working alone is 1.7 times more than efficient than a team consisting of 2--10 people.
Furthermore, we can see that the efficiency of the members decreases significantly when there are more than 60 members, while the efficiency from 2 and 60 members is roughly constant.
Thus, we conclude that it is undesirable to involve more than 60 developers in a project if we want the project members to work efficiently.

%
%
\subsection{Workload Bias between Internal Members}

\begin{figure*}[t]
  \begin{center}
  \begin{minipage}{0.32\hsize}
    \begin{center}
      \includegraphics[width=60mm]{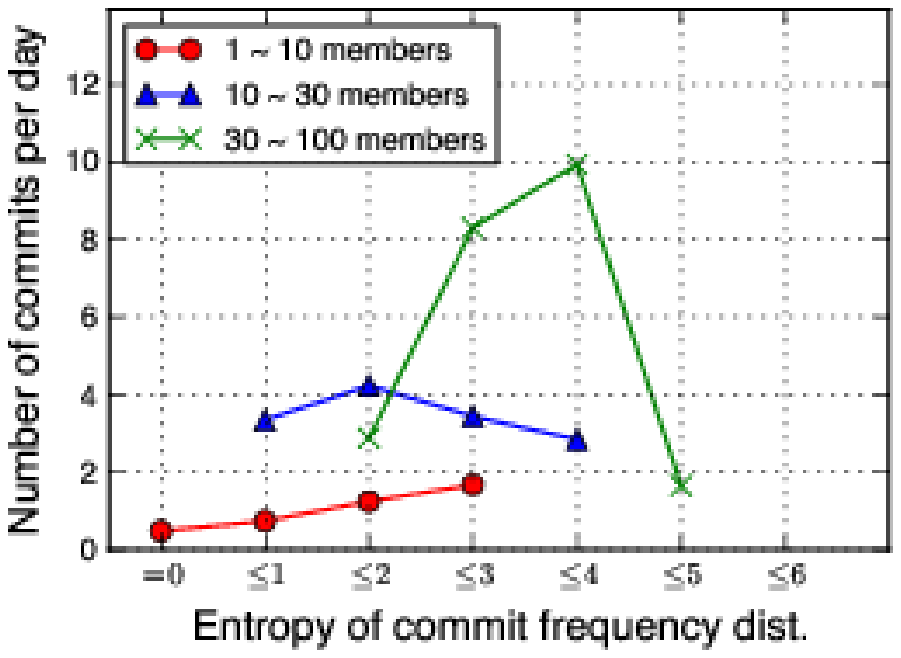}\\
      \vspace{-1mm}      
      (a) $\mathtt{Commit}$
    \end{center}
  \end{minipage}
  \begin{minipage}{0.32\hsize}
    \begin{center}
      \includegraphics[width=60mm]{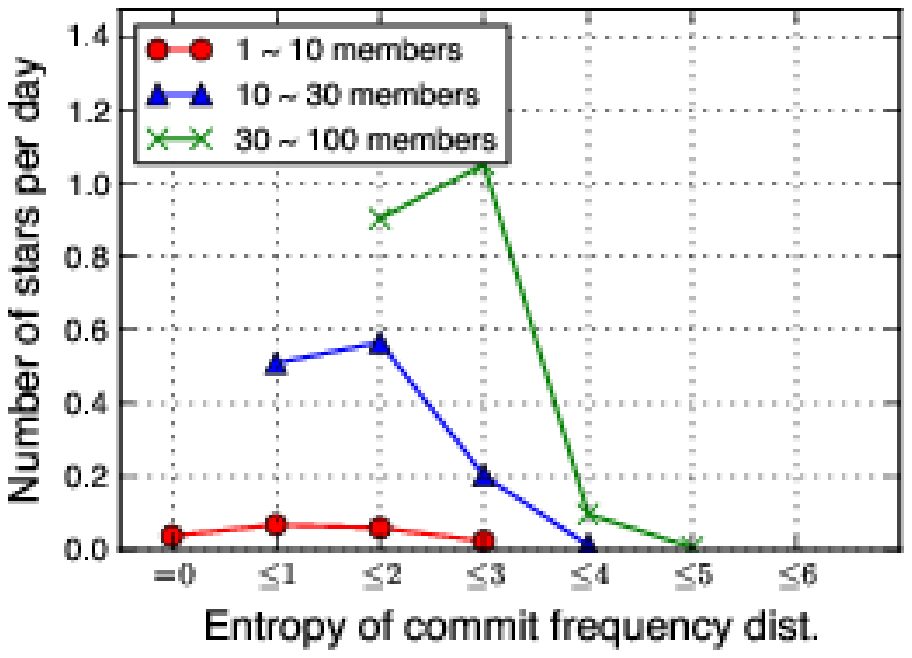}\\
      \vspace{-1mm}      
      (b) $\mathtt{Star}$
    \end{center}
  \end{minipage}
  \begin{minipage}{0.32\hsize}
    \begin{center}
      \includegraphics[width=60mm]{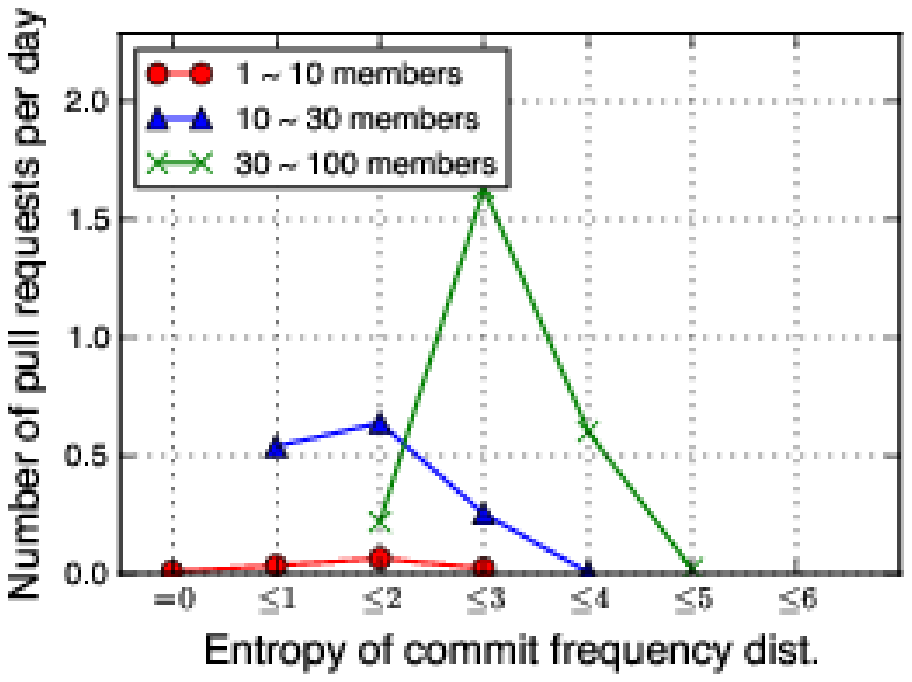}\\
      \vspace{-1mm}     
      (c) $\mathtt{PullReq}$
    \end{center}
  \end{minipage}
 \vspace{1mm}
 \caption{The relationships between workload bias and each of the success indexes. Red circles, blue triangles and green crosses indicate projects with 1--10, 10--30 and 30--100 members, respectively.}
 \label{fig:workbias_vs_success}
 \end{center}
\end{figure*}

We next analyze the relationship between workload bias among internal members and the success indexes.
For workload bias, we consider two extreme cases:
\begin{enumerate}
\item Only a few internal members work hard.
\item All the internal members work uniformly hard.
\end{enumerate}
Our question relates to which case projects achieve success in social coding.
To quantify the difference between the cases, we define $\mathtt{WorkBias}$ as the entropy of the commit distribution over internal members, as follows:
\begin{equation}
\mathtt{WorkBias} = -\sum_{m \in M} \theta_m \log \theta_m
\label{eq:workbias}
\end{equation}
where, $M$ represents a set of internal members of a given project and $\theta_m$ represents the ratio of commits performed by member $m$ in the project.
$\mathtt{WorkBias}$ approaches zero when the commits of the project concentrate on only a few members but has a high value when the commits are evenly distributed to all the members.

Figure~\ref{fig:workbias_vs_success} shows the relationships between the workload bias and the success indexes.
As shown in Figure~\ref{fig:workbias_vs_success}(a), which shows the effects on $\mathtt{Commit}$, when the number of internal members is between 1 and 10, projects with a higher workload bias have a higher $\mathtt{Commit}$.
On the other hand, with 10--30 members and 30--100 members, projects with a higher $\mathtt{WorkBias}$ have a lower $\mathtt{Commit}$ and projects distributing work moderately have the highest $\mathtt{Commit}$.
Interestingly, we find that $\mathtt{WorkBias}$ relates to $\mathtt{Star}$ and $\mathtt{PullReq}$, which represent evaluations from external developers.
In the results of both $\mathtt{Star}$ and $\mathtt{PullReq}$, we can find that lower $\mathtt{WorkBias}$ is better than higher one, except for 30--100 members on $\mathtt{PullReq}$.
We also find that a moderate $\mathtt{WorkBias}$ has the highest $\mathtt{Star}$ and $\mathtt{PullReq}$ as with the result for $\mathtt{Commit}$.

%
%
%
%

\begin{table}[t]
\caption{Features representing response to external developers.}
\vspace{2mm}
\centering
\begin{tabular}{rp{13em}}
\hline
Feature & Description \\
\hline
$\mathtt{ResponseRate}$ & Ratio of internal members who responded to pull requests within a month. \\
$\mathtt{MergeRate}$ & Ratio of pull requests that were brought in. \\
$\mathtt{ResponseTime}$ & Average time between arrival of pull requests and response to them.  \\
$\mathtt{PullReqByInside}$ & Number of pull requests from internal members. \\
\hline
\end{tabular}
\label{tab:social_features}
\end{table}

\begin{table}[t]
  \centering
  \caption{Kendall rank correlation coefficient between the features representing the response to external developers and the success of projects. All the elements exhibit statistically significant correlations (t-test, $p < 0.01$).}
  \vspace{2mm}
    \begin{tabular}{r|ccc}
    \hline
          & $\mathtt{Commit}$ & $\mathtt{Star}$ & $\mathtt{PullReq}$ \\
    \hline
    $\mathtt{ResponseRate}$ & 0.01  & 0.39  & 0.83 \\
    $\mathtt{MergeRate}$ & 0.09  & 0.37 & 0.68 \\    
    $\mathtt{ResponseTime}$ & -0.04 & 0.20 & 0.14 \\
    $\mathtt{PullReqByInside}$ & 0.14  & 0.27  & 0.31 \\
    \hline
    $\mathtt{Commit}$ & 1.00  & 0.06  & 0.03 \\    
    \hline
    \end{tabular}%
  \label{tab:corr_social}%
\end{table}%

%
%

%
%

\section{Effect of Response to External Developers}
A key function in social coding is that external developers can contribute to projects other than projects they have already joined via {\it pull request}.
A pull request is usually used for reporting bugs and improvements and requesting the addition of new functions.
Thus, it is expected that utilizing the pull requests from external developers meaningfully leads the projects developing improved software.
We then ask the question:
how do successful projects utilize and respond to pull requests?
To answer this question, we analyze the relationships between the response to the pull requests and the success of project.

As in Section \ref{sec:member}, we perform a correlation analysis to investigate the effects of the response to the pull requests.
To accomplish this, we first extract features representing the response to external developers.
Table~\ref{tab:social_features} lists their features and descriptions.

Table \ref{tab:corr_social} shows the Kendall rank correlation coefficients (indicated by $\tau$) between the features representing the response to pull requests and the success indexes.
First, we find that $\mathtt{ResponseRate}$ is strongly correlated with $\mathtt{Star}$ ($\tau=0.39$) and $\mathtt{PullReq}$ ($\tau=0.83$) but not with $\mathtt{Commit}$ ($\tau=0.01$).
Moreover, we also find that $\mathtt{MergeRate}$ is strongly correlated with $\mathtt{Star}$ ($\tau=0.37$) and $\mathtt{PullReq}$ ($\tau=0.68$).
These results indicate that projects which tackle the pull requests faithfully achieve success in terms of popularity and sociality.
Since $\mathtt{ResponseRate}$ represents the ratio at which internal members responded to pull requests within a month, a speedy response seem to be important.
Interestingly, $\mathtt{ResponseTime}$ is positively correlated with $\mathtt{Star}$ ($\tau=0.20$) and $\mathtt{PullReq}$ ($\tau=0.14$).
This means that the longer it takes to respond to the pull requests the better.
According to these results, it is important to respond to the pull requests faithfully without worrying about the response speed.

The pull requests can be sent from internal members with the advantage that the software changes are performed in a social environment.
Since external developers can join in the middle of their development, it is important to ensure the transparency of the code changes.
In fact, $\mathtt{PullReqByInside}$ is correlated with $\mathtt{Star}$ ($\tau=0.27$) and $\mathtt{PullReq}$ ($\tau=0.31$), where we note $\mathtt{PullReq}$ is calculated based only on pull requests from external developers.
According to the fact that $\mathtt{Commit}$ is poorly correlated with $\mathtt{Star}$ ($\tau=0.06$) and $\mathtt{PullReq}$ ($\tau=0.03$), we can say that proceeding with development via the pull requests is more likely to lead to success.

%
%
%
%
\section{Effect of Software Developed by Projects}
Lastly, we analyze the effects of software developed by projects on the success of project.
Unfortunately, GitHub does not provide category and tag information that represent the characteristics of the software.
We thus attempt to characterize the software based on the contents of README files.

%
%
%

%
\begin{figure*}[t]
  \begin{center}
  \begin{minipage}{0.32\hsize}
    \begin{center}
      \includegraphics[width=55mm]{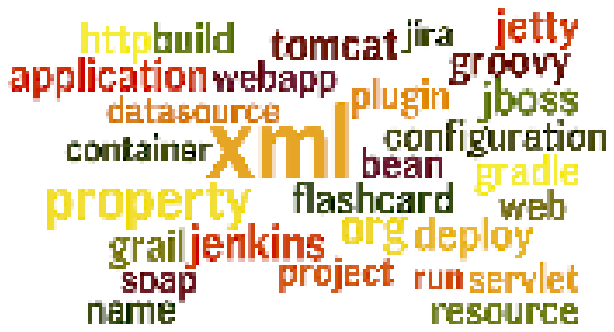}\\
      \vspace{2mm}
      (a) $\mathtt{Commit}$
    \end{center}
  \end{minipage}
  \begin{minipage}{0.32\hsize}
    \begin{center}
      \includegraphics[width=52mm]{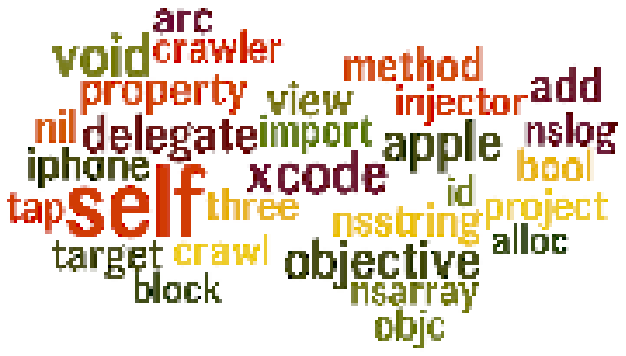}\\
      \vspace{2mm}
      (b) $\mathtt{Star}$ 
    \end{center}
  \end{minipage}
  \begin{minipage}{0.32\hsize}
    \begin{center}
      \includegraphics[width=58mm]{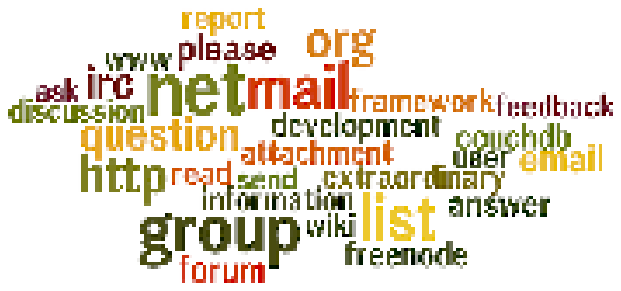}\\
      \vspace{3.5mm}
      (c) $\mathtt{PullReq}$
    \end{center}
  \end{minipage}
 \vspace{1mm}
 \caption{Words in topics with the highest coefficients in the regression models. The font size reflects the word distribution of each topic.}
 \label{fig:tagcloud}
 \end{center}
\end{figure*}
%

%
%
%
README usually describes the description of projects and their softwares usage.
In GitHub, this is described at the front page of each project as shown in Figure~\ref{fig:github} and is necessary that external developers understand what the project addresses.
To understand what successful projects describe in README, we analyze relationship between the content of README and success.

\subsubsection{Topic Extraction}
To understand the characteristics of software from README automatically, we run Latent Dirichlet Allocation (LDA)~\cite{Blei2003}, which is a method for jointly learning the word distributions of topics and the topic distributions of documents, for the contexts of README.
We therefore learn linear regression models with feature selection, which are called Lasso~\cite{Tibshirani1996}, using the learned topic distribution of each README to investigate what topics affect the success indexes.
In our study, we set the number of topics on LDA at 80 and the regularization parameter $\alpha$ of Lasso at $10^{-4}$.
By combining LDA and Lasso, it is expected that, if a topic contributes positively to the success indexes, then the coefficient of the topic is positive, and otherwise negative.

Figure~\ref{fig:tagcloud} shows words in topics with the highest coefficients in the regression models for $\mathtt{Commit}$, $\mathtt{Star}$ and $\mathtt{PullReq}$, respectively.
As shown in the result of $\mathtt{Commit}$, we can find that the topic includes words specific to web applications such as webapp, web, servlet, tomcat and grail. 
Moreover, words such as configuration, build and run often indicate how to install command line tools.

In the $\mathtt{Star}$ result, we can find words indicating software for iOS and MacOS such as apple, iphone, objc and xcode.
This result
indicates that projects developing software running on Apple's products attain popularity.
Unlike the above two results, the result of $\mathtt{PullReq}$ consists of words representing communication such as mail, ask, discussion and question, rather than information about software.
This means that telling external developers of projects the communication methods to leads to receive many pull requests.

%
%
%
%
\section{Related Work}\label{sec:related_work}
\subsection{Studies of Social Coding}
To understand structures and developer behaviors in social coding, a few studies have analyzed data crawled from GitHub.
For example, Takhteyev et al. investigated the geographical distance between developers who were following each other or who had joined the same projects by estimating the geography of the developers from the noisy location description~\cite{Takhteyev2010}.
Other studies addressed the problems of finding popular developers and projects and recommending projects.
Thung et al. proposed a method for discovering famous developers and projects by applying the PageRank algorithm to a developer-developer network and project-project network~\cite{Thung2013}.
Sarma et al. proposed a graph-based developer recommendation system by extending the SimRank algorithm~\cite{Sarma2012}.
Orii applied the Collaborative Topic Model~\cite{Wang2011} to recommend projects on GitHub to developers~\cite{Orii2012}.
Recently, Lima et al. analyzed the events happening on GitHub based on network analysis.~\cite{Lima2014}.

In software engineering, Allamanis et al. showed that the performance of code suggestion improves by learning a language model using massive source codes in GitHub repositories~\cite{Allamanis2013}.
Pham et al. performed interviews and questionnaires for GitHub users and investigated how the transparency found on GitHub influences developers' testing behaviors~\cite{Pham2013}.
To predict future software failure caused by large collaboration on GitHub, Ell proposed a method finding developer pairs that  would cause failed code changes based on technical networks~\cite{Ell2013}.

However, there is no study that focuses on project team structures and developer behaviors on GitHub.
To the best of our knowledge, our study is the first to analyze the characteristics of successful projects in social coding from various perspectives.

\subsection{Studies of Project Success}
For decades, project success has been well studied in business economics~\cite{Pinto1990,Raymond1998,Wit1988,Munns1996,Might1985} and software engineering~\cite{Thamhain1987,Deutsch1991,Grewal2006}.
Recently, various projects and teams such as research projects, sport teams and musical groups are analyzed for understanding the topological characteristics in their collaboration networks~\cite{Guimera2005,Duch2010,Uzzi2008}.


%

%
Before the late '90s, many of studies in software engineering focused on closed projects~\cite{Thamhain1987,Deutsch1991,Larson1989}.
After {\it Open Source} attracted attention, SourceForge.net~\footnote{\url{http://sourceforge.net/}} has became famous as a web service providing a place to release open source software.
The emergence of this web service has meant that researchers are able to perform an analysis using tens of thousands of projects.
There are studies that investigate the success of open source projects using data collected from SourceForge.net.
To forecast and investigate the success of open source projects, some studies defined different project success measures such as the number of developers and subscribers~\cite{Sen2012} (corresponding to $\mathtt{Star}$), the amount of developer activity~\cite{Subramaniam2009} and the time taken to fix software bugs~\cite{Crowston2006}, 
There are also qualitative measures such as code quality and opinions on mailing lists~\cite{Crowston2003}.

The SCSs are partially similar to SourceForge.net.
In fact, Peterson reported that, in GitHub, many of the traditional aspects of open source software development remain, such as the fact that most project development is undertaken by a small group of core developers~\cite{Peterson}.
Nevertheless, there are two obvious differences between the SCSs and SourceForge.net.
The first is that the projects in the SCSs open not only the software source codes but also the development process.
Namely, it means that the external developers of the projects can easily join the development.
The second is that the projects are on a developers' social network.
The developers can monitor the activities of developers who they follow such as project making, contribution to projects and bookmarks. 
Thus, the projects can be shared among developers by propagating the activities through the social network.
This phenomenon would encourage collaboration between developers with the same interests and purpose. 
We believe that our work is the first step toward revealing the differences between the SCSs and SourceForge.net as well as understanding the impact of these differences on collaboration.

%
%
%
%
\section{Conclusion}
In this paper, we focused on social coding, which is a new form of software development on the Web, and investigated the differences between successful and unsuccessful projects.
To evaluate the success of a project quantitatively, we defined three measures as success indexes: activity, popularity and sociality.
Activity is calculated based on the update frequency of the content developed by each project.
Popularity and sociality are evaluated by the number of bookmarks and the number of updates by the external developers of projects.
The analysis was performed based on three perspectives:
1) team structure, 2) social activity with external developers, and 3) content developed by projects.
Through the analysis, we found the following relationships between the success indexes and the characteristics of projects:
\begin{itemize}
\item Projects with larger numbers of internal members have higher activity, popularity and sociality. However, the work effectiveness of each developer decreases significantly when the number of internal members exceeds 60.
\item Projects in which the internal members are well connected but the distance between the members is considerable are more likely to exhibit higher activity, popularity and sociality.
\item If the number of internal members is fewer than 10, projects which evenly distribute work between members are more likely to have higher activity. If there are more than 30 members, the activity is high when the work is moderately distributed.
\item Projects that faithfully tackle the change requests from external developers are more likely to exhibit higher popularity and sociality.
\item Improving the transparency of software changes, e.g., the internal members proceed with projects via pull requests, would lead to higher popularity and sociality.
\item Projects for web application development would have high activity. Also, software running on Apple's products would lead to high popularity.
\item Projects stating the communication methods with members inside the projects have higher sociality.
\end{itemize}

Our work is the first step toward understanding how to proceed with projects in online social environments.
We expect a deeper knowledge of project success to be acquired by using our datasets.




\bibliographystyle{aaai}
\bibliography{yuyay-icwsm2014}

\end{document}